%
\catcode`@=11 
\font\sevensy=cmsy7
\font\titlefontB=cmssdc10 at 18pt
\font\sect=cmssdc10 at 12pt
\font\tmsm= zptmcm7t at 14pt
\font\petit=msbm8
\font\d= msbm10
%
 
%

%

%

%
\catcode`@=12 
\def\sot#1\over#2{%
\mathrel{ \mathop{#2}\limits_{#1}}}
%
%
%
%

%
%
%
\def\downnormalfill{$\,\,\vrule depth4pt width0.4pt
\leaders\vrule depth 0pt height0.4pt\hfill\vrule depth4pt width0.4pt\,\,$}
\def\WT#1{\mathop{\vbox{\ialign{##\crcr\noalign{\kern3pt}
      \downnormalfill\crcr\noalign{\kern0.8pt\nointerlineskip}
      $\hfil\displaystyle{#1}\hfil$\crcr}}}\limits}

%
%
%
%
%
%
\magnification = \magstep1
\hsize = 16.9 truecm
\vsize = 21.6 truecm
\parindent=0.9 truecm
\parskip = 2pt
\normalbaselineskip = 13pt plus 0.2pt minus 0.1pt
\baselineskip = \normalbaselineskip
\topskip= 17pt plus2pt      
\voffset= 0 truecm   
\hoffset= -0.01 truecm      
%
%
%
%
\def\tfract#1/#2{{\textstyle{\raise0.8pt\hbox{$\scriptstyle#1$}\over%
\hbox{\lower0.8pt\hbox{$\scriptstyle#2$}}}}}

\def\radi2k{\tfract 1/{\sqrt {2k}} }
\def\der{\partial }
\headline={\ifnum\pageno=0\line{ }\else {\vbox{\line{\hfil {{\rm \folio}}}\line{\hrulefill}}}\fi}
%
\def\go{\leavevmode \raise.3ex\hbox{$\scriptscriptstyle\langle\!\langle\, $}%
~\ignorespaces}

\def\gf{\relax \ifhmode \unskip~\else \leavevmode \fi \raise.3ex\hbox{$\scriptscriptstyle\rangle\!\rangle\, $}}
%

\newbox\novebox
\setbox\novebox=\hbox{\kern+.3em {$V_t$}%
\kern-0.79em\raise1.8ex\hbox{\sevensy {\char'016}}\kern+.3em}

\newbox\ottobox
\setbox\ottobox=\hbox{\kern+.3em {$V$}%
\kern-0.61em\raise1.8ex\hbox{\sevensy {\char'016}}\kern+.3em}
\def\vnt{\copy\ottobox}

\input graphicx
%
%
\def\putdim#1#2#3{\ifvmode\vbox to0pt{\kern-#2\hbox to 0pt{\kern#1{#3}\hss}%
\vss}\nointerlineskip\else\rlap{\kern#1\raise#2\hbox{#3}}\fi}%
\def\hangmore#1#2{\noindent\hangindent-#2\hangafter-#1}%
\def\leftinsert#1#2{\setbox92=\hbox{L}%
\count99=\ht91 \advance\count99 by \dp91%
\divide\count99 by\baselineskip \advance\count99 by 2%
\vskip0pt\putdim{#1}{0pt}{\box91}\vskip-\ht92%
\hangmore{\count99}{#2}}%
%
%
%
%
%
%

\nopagenumbers
\count0=0

\null

\line{\hfil IFUP-TH/2013-03}
\line{\hfil LAPTH-003/13}

\vskip 2.00 truecm

\centerline {\titlefontB Three-manifold invariant}

\vskip 0.4 truecm

\centerline {\titlefontB from functional integration}

\vskip 2.5 truecm

\centerline {\tmsm Enore Guadagnini$^{\, a}$ and Frank Thuillier$^{\, b}$}

\vskip 1.0 truecm

\centerline{\sl $^{a}$ Dipartimento di Fisica ``E. Fermi" dell'Universit\`a di Pisa
 and INFN, Sezione di Pisa, Italy.}
\centerline{\sl $^b$ LAPTH, Chemin de Bellevue, BP 110, F-74941 Annecy-le-Vieux cedex, France.}

\vskip 4.2 truecm

\centerline {\bf Abstract}

\medskip

\midinsert \narrower
We give a precise definition and produce a path-integral computation of  the normalized partition function of the abelian $U(1)$ Chern-Simons field theory defined in a general closed oriented 3-manifold. We use the Deligne-Beilinson formalism, we  sum over the inequivalent $U(1)$ principal bundles over the manifold and, for each bundle, we integrate  over the gauge orbits of the associated connection 1-forms. The result of the functional integration is  compared  with the abelian $U(1)$ Reshetikhin-Turaev surgery invariant.

\endinsert

\vskip 1 truecm


\vfill\eject

\noindent  {\sect  1. Introduction}

\vskip 0.4 truecm

Gauge quantum field theories play a fundamental role in the description of physical phenomena. Most of the models  that have been considered so far  are defined in Minkowski space. But one can imagine that, in certain conditions, it will become important to study a quantum gauge theory defined in a topological nontrivial manifold.   In this paper we will consider a quantum field theory with a  $U(1)$ local gauge symmetry which is defined  in a general connected closed oriented 3-manifold $M$; the action is given by the Chern-Simons functional and the observables of this model represent topological invariants [1,2,3,4,5,6].

In the present article we shall concentrate on the quantum field theory aspects which are related with the path-integral definition and with the computation of the normalized partition function of the theory, which represents a topological invariant of the 3-manifold $M$. By using the Deligne-Beilinson formalism,  it turns out that  the result of the functional integration for the normalized partition function of the  U(1) Chern-Simons  theory is strictly related with the Reshetikhin-Turaev  $U(1)$ surgery invariants of 3-manifolds [7,8,9].

\vskip 0.9 truecm

\noindent {\sect 1.1 Summary and results}

\vskip 0.4 truecm

Let us give a short description of the content of our paper and a presentation of  the main results.  In the Deligne-Beilinson (DB) formalism [5,10,11,12,13], each gauge orbit $A$ of a $U(1)$ connection on the 3-manifold $M$ is a class belonging to  the DB cohomology space $H^1_D(M)$. The Chern-Simons action $S$ is given [5] by
$$
S [A] = 2 \pi k \int_M A * A \; ,
\eqno(1.1)
$$
where  the $*$-product denotes  the pairing  $H^1_D (M) \times H^1_D (M) \rightarrow H^3_D (M)\sim \hbox{\d R}/ \hbox{\d Z}$ which is associated with the canonical DB product [13].  A modification of the orientation of the manifold $M$ is equivalent to a change in the sign of the integer coupling constant $k$, so one can choose  $k > 0 $. Let $\Omega^1 (M) $ be the space of the 1-forms on $M$ and
$\Omega^1_{\hbox{\petit Z}} (M) $
 the subspace of closed forms with integral periods. The space $\Omega^1_{\hbox{\petit Z}} (M) $ corresponds to the set of gauge transformations. A presentation of $H_D^1(M)$ is given [10,11,12,13,14] by the following exact sequence
$$
0 \rightarrow \Omega^1(M) / \Omega^1_{\hbox{\petit Z}} (M) \rightarrow H^1_D(M) \rightarrow H^2(M) \rightarrow 0 \; ,
\eqno(1.2)
$$
in which $H^2 (M) $ denotes the second integral cohomology group of $M$ and,  because of  Poincar\'e duality,
$H^2 (M) \simeq H_1(M)$ where $H_1(M)$ stands for the  first homology group of $M$.    Thus  $H_D^1(M)$ can be understood as an affine bundle  over $H_1(M)$  for which   $\Omega^1(M) / \Omega^1_{\hbox{\petit Z}} (M)$ acts as a translation group on the fibres. More precisely, each fibre is characterized by an element of $H_1(M)$; a generic DB class $A$ that belongs to the fibre over $\gamma \in H_1(M)$  can be written as
$$
A = \widehat A_{\gamma} + \omega   \; ,
\eqno(1.3)
$$
where    $ \omega   \in \Omega^1(M) / \Omega^1_{\hbox{\petit Z}} (M)$. The  element $\widehat A_{\gamma}$ just fixes an origin on the fibre over $\gamma$ and any other element of this fibre can be obtained from $\widehat A_{\gamma}$ by means of a translation  with the 1-form $\omega $ modulo closed forms of integral  periods. For each fibre, the choice of the corresponding origin class $\widehat A_{\gamma}$ is not unique. On the fibre over the trivial element of $H_1(M)$ one can take as  canonical origin the zero class,
$\widehat A_0 =0 $, which is precisely the gauge orbit of the vanishing connection.

Each DB class $A \in H^1_D(M)$ describes a $U(1)$ principal bundle with connection (up to gauge transformations), and equation (1.2) shows that the inequivalent principal $U(1)$ bundles can be labelled by  $H_1(M)$. Let us assume [5] that the functional integration consists of a sum over the inequivalent principal bundles and, for each bundle, of a sum  over the  gauge orbits of the corresponding connection 1-forms. According to  equation (1.3), this means  that the path-integral is given by $\int D A \, e^{i S[A] } = \sum_{\gamma \in H_1(M)}\int D \omega \, e^{i S [\widehat A_{\gamma} + \omega] }$  where one has a sum over all the elements  of the homology group of the manifold.

Let us  define the normalized partition function $Z_k(M)$ as
$$
Z_k(M) = {\sum_{\gamma \in H_1(M)} \int D \omega \; e^{i S[\, \widehat A_{\gamma} + \omega ] } \over \int D \omega \;  e^{i S[ \omega ] } } \; .
\eqno(1.4)
$$
The normalization factor  $\int D \omega \, e^{i S[ \omega]} = \int D \omega \, e^{i S[\, \widehat A_0 + \omega]}$ just   corresponds  to the functional integral associated with the fibre of $H^1_D(M)$ over the trivial element $0 \in H_1(M)$; i.e. $\int D \omega \, e^{i S[ \omega]} $ represents the integral over the gauge orbits of the connection  1-forms  of the trivial principal bundle over $M$.

\noindent {\bf Remark 1.1.} In quantum field theories one is really concerned with distributional fields, so one may be  interested in the possible modifications of sequence (1.2) under rough extensions of the fields space.   Quite remarkably, the basic structure of the configuration space ---as described by sequence (1.2)--- is stable under the inclusion of distributional configurations. Indeed there is a natural inclusion  [5,6] of $H^1_D(M)$ and of the space $Z_1(M)$ of 1-cycles in $M$ into the Pontrjagyn dual $Hom (H^1_D(M), S^1)$ of $H_D^1(M)$. These inclusions are ensured by the canonical DB product and the $\hbox{\d R}/ \hbox{\d Z}$-valued integration over 1-cycles of $M$. This dual space contains generalized (i.e. distributional)  connections and it is  embedded into the  exact sequence
$$
0 \rightarrow Hom ( \Omega^2_{\hbox{\petit Z}}(M), S^1 ) \rightarrow Hom (H^1_D(M), S^1 ) \rightarrow H^2(M) \rightarrow 0 \; .
\eqno(1.5)
$$
Let us introduce the simplified notation
$$
H^1_D(M)^* \equiv Hom (H^1_D(M), S^1 ) \quad , \quad \Omega^2_{\hbox {\petit Z}} (M)^* \equiv Hom ( \Omega^2_{\hbox{\petit Z}}(M), S^1 ) \; ;
\eqno(1.6)
$$
note that there also is a natural inclusion
$$
{\Omega^1 (M) \over \Omega^1_{\hbox{\petit Z}}(M)} \hookrightarrow \Omega^2_{\hbox {\petit Z}} (M)^* \; .
\eqno(1.7)
$$
Equation (1.3) admits a   distributional extension in which $ \widehat A_\gamma \in H^1_D(M)^* $ and $\omega \in \Omega^2_{\hbox {\petit Z}} (M)^* $.

In general,  the abelian homology group $H_1(M)$ can be decomposed as $\, H_1(M) = F(M)  \oplus T(M)$, where $F(M)$ is  freely generated   and the torsion component $T(M)$ can be written as a direct sum of $\hbox{\d Z}_p \equiv \hbox{\d Z}/ p  \hbox{\d Z}$ factors. For torsion-free manifolds, when  the torsion component $T(M)$ is trivial, the main properties of the path-integral have been studied in Ref.[5]. In the present article we shall concentrate on the pure torsion case,  in which  the freely generated component  $F(M)$ is trivial and then $H_1(M)$ is a finite group
$$
H_1(M) = T(M) = \hbox{\d Z}_{p_1} \oplus  \hbox{\d Z}_{p_2} \oplus \cdots \oplus  \hbox{\d Z}_{p_w} \; ,
\eqno(1.8)
$$
in which the   torsion numbers $\{ p_1 , p_2 ,..., p_w \} $ are fixed by the convention that  $p_i $ divides $p_{i+1}$.
Some preliminary results on the pure torsion case have been discussed in Ref.[6].  The action (1.1) is a quadratic function of the fields and then the result of the functional integration (1.4) does not depend on the particular choice of the origin class $  \widehat A_{\gamma} $ for each $\gamma \in H_1(M)$.

\vskip 0.3 truecm

\noindent {\bf Proposition 1.} {\it For each torsion element $\gamma \in T(M)$,  one can  select the  origin class $\widehat A_{\gamma} $ to correspond to a  stationary point of the action, i.e. $\widehat A_{\gamma} $ can be chosen to be equal to  the gauge orbit $  A^0_{\gamma} $ of a flat connection.  Therefore   the normalized partition function} (1.4) {\it can be written as a sum over the gauge orbits of flat connections }
$$
Z_k(M) = {\sum_{\gamma \in H_1(M)} \int D \omega \; e^{i S[A_{\gamma}^0 + \omega]  }  \over \int D \omega \;  e^{i S[ \omega ] } } = \sum_{\gamma \in H_1(M)} e^{i S [   A^0_{\gamma} \, ] } \; .
\eqno(1.9)
$$

\vskip 0.3 truecm

Indeed,  $S[A_{\gamma}^0 + \omega]  = S[A_{\gamma}^0] + S[\omega ] + 2 \pi k \int A_{\gamma}^0 * \omega$ but since $A_{\gamma}^0$ is the class of a flat connection and $\omega $ is globally well defined in $M$, the last term  is vanishing and therefore $S[A_{\gamma}^0 + \omega]  = S[A_{\gamma}^0] + S[\omega ] $. Consequently
$$\eqalign {
Z_k(M) &= {\sum_{\gamma \in H_1(M)} \int D \omega \; e^{i S[A_{\gamma}^0] } \, e^{ i S[ \omega]  }  \over \int D \omega \;  e^{i S[ \omega ] } } = \cr
&= \sum_{\gamma \in H_1(M)} e^{i S[A_{\gamma}^0] } { \int D \omega \; e^{ i S[ \omega]  }  \over \int D \omega \;  e^{i S[ \omega ] } } =
\sum_{\gamma \in H_1(M)} e^{i S [   A^0_{\gamma} \, ] } \; . \cr }
\eqno(1.10)
$$
On the other hand, since the value of the path-integral does not depend on the choice of the origins in $H^1_D(M)^*$,   for each $\gamma$ one finds
$$
e^{i S [   A^0_{\gamma} \, ] } = { \int D \omega \; e^{i S[\, \widehat A_{\gamma} + \omega ] } \over \int D \omega \;  e^{i S[ \omega ] } } \; .
\eqno(1.11)
$$
If $\widehat A_{\gamma}$ satisfies $S[\, \widehat A_{\gamma} ] =0 \; \;  \hbox{mod {\d Z}}$,   then
$$
e^{i S [   A^0_{\gamma} \, ] } = { \int D \omega \; e^{i S[ \omega ] } e^{4 \pi i k \int \omega * \widehat A_{\gamma}} \over \int D \omega \;  e^{i S[ \omega ] } } \; ,
\eqno(1.12)
$$
and by means of the path-integral (1.12) one can compute the amplitude  $e^{i S [   A^0_{\gamma} \, ] }$.

Let us introduce a set of generators
$\{ h_1, h_2,..., h_w \}$ for $H_1(M)$; the element $h_i$ is a generator for $\hbox{\d Z}_{p_i} $, with $p_i h_i = 0 $.
A generic element  $\gamma \in H_1(M) $ can be described by means of the sum $ \gamma = \sum_{i=1}^w n_i h_i $ with  integers $\{ n_i \} $. Each  term $e^{i S [   A^0_{\gamma} \, ] }$ can now be written as
$$
e^{i S [   A^0_{\gamma} \, ] } = e^{  2 \pi i k \sum_{ij} n_i n_j Q_{ij}  }  \; ,
\eqno(1.13)
$$
where the matrix $Q_{ij}$ determines   a  $\hbox{\d Q} / \hbox{\d Z}$-valued quadratic form $Q$ on the torsion group $T(M)$.
Although $Q$ only depends [16,17,18] on the manifold $M$, in order to describe the result of the functional integration (1.12),  it is useful to  consider a surgery presentation  [19] of  $M$ in $S^3$.

  Let ${\cal L} = {\cal L}_1 \cup {\cal L}_2 \cdots \cup {\cal L}_m \subset S^3 $ be a framed surgery link ---associated with  a Dehn surgery presentation   of $M$ in $S^3$---  with integer surgery coefficients and let  $\hbox{\d L}$ denote the corresponding linking matrix. When the homology group is given by equation (1.8), one can always find a surgery presentation  in which the linking matrix $\hbox{\d L}$ is non-degenerate (invertible), so we assume that this is indeed the case.
  For each link component ${\cal L}_t$ (with $t = 1,2,...,m$),  let $G_t$ be a simple small circle linked with ${\cal L}_t$ which can be taken as a generator of the homology of the complement of ${\cal L}_t$ in $S^3$; then $\{ G_1 ,..., G_m \}$ is a set of generators  for the homology of $S^3 - {\cal L}$. The homology group $H_1(M)$ admits the presentation
$$
H_1(M ) = \langle G_1,..., G_m \, | \, [ {\cal L}_{1 \rm f}] =0 , [ {\cal L}_{2 \rm f}] =0 ,..., [ {\cal L}_{m \rm f}]  =0 \, \rangle
\; ,
\eqno(1.14)
$$
where $[ {\cal L}_{t \rm f}] $ is the homology class (in $S^3 - {\cal L}$) of the framing ${\cal L}_{t \rm f}$ of the component ${\cal L}_t$
$$
[{\cal L}_{t \rm f}] = \sum_{s=1}^m \,  \hbox{\d L}_{ts} \, G_s \; .
\eqno(1.15)
$$
 Each generator $h_i$ of $H_1(M)$ can be written as a linear combination  of the $\{ G_t \} $ generators
 $$
 h_i = \sum_{t=1}^m B_{it} G_t \quad , \quad ( i = 1,2..., w ) \; ,
 \eqno(1.16)
 $$
 with integer  coefficients $B_{it}$.

\vskip 0.3 truecm

\noindent {\bf Corollary 1.} {\it  In the basis defined by the generators $\{ h_i \}$,  the matrix elements $Q_{ij}$ of the quadratic  form  on  the torsion group $T(M)$ are given by }
$$
Q_{ij} = \sum_{t,s=1}^m \, B_{it} \, B_{js} \, \hbox{\d L}^{-1}_{ts} \; ,
\eqno(1.16)
$$
{\it where $\hbox{\d L}^{-1}$ represents the inverse in $\hbox{\d R}^m\!$ of the linking matrix;   the normalized partition function $Z_k(M)$ takes the form
$$
Z_k(M)  = \sum_{n_1=0}^{p_1-1} \sum_{n_2=0}^{p_2-1}\cdots  \sum_{n_w=0}^{p_w-1}e^{ 2 \pi i k \sum_{ij} n_i n_j Q_{ij}  }  \; .
\eqno(1.17)
$$
 Moreover,
$$
Z_k(M) = \left ( p_1 p_2 \cdots p_w \right )^{1/2}\,  I_k (M) \; ,
\eqno(1.18)
$$
where $I_k(M)$ denotes the value of the  Reshetikhin-Turaev  $U(1)$ surgery invariant  of the  3-manifold $M$. }

\vskip 0.3 truecm

As a matter of facts, in the definition and in the computation of the normalized partition function $Z_k(M)$ of the $U(1)$ Chern-Simons theory there is no need of introducing  a metric in the 3-manifold $M$; moreover, $Z_k(M)$  has nothing to do with the perturbative gauge-fixing procedure.

The paper is organized as follows. The basic rules which are used in the computation of the field theory path-integrals are listed in Section~2. Section~3 contains a proof of Proposition~1 and Corollary~1 together with a path-integral derivation of expression (1.17) of the normalized partition function. One illustrative example is presented in Section~4.

\vskip 1.2 truecm

\noindent {\sect 2. Computation rules}

\vskip 0.4 truecm

In certain expressions of the previous section, ratios of functional  integrations ---as indicated for instance  in equations (1.4) and (1.12)---  appear. This notation belongs to the set of standard  conventions which are employed in physics, in which any meaningful quantity  takes the form of a ratio of regularized functional integrations in the limit in which the regularization is removed.

\noindent {\bf Remark 2.1.} Each functional integration, which  formally involves an infinite number of integration variables, can be approximated or regularized by restricting the integral to a finite number $N$ of variables;  this regularization is removed in the $N \rightarrow \infty $ limit. The ratio of two path-integrals means: (1) introduce a regularization in the numerator and in the denominator simultaneously (with the same finite $N$), (2) for each fixed $N$, the regularized ratio ---that is the ratio of the two regularized integrals---  is well defined and depends on $N$,    (3) finally  consider  the $N\rightarrow \infty $ limit of the regularized ratio.  For the ratios of functional integrations  considered in quantum field theory, this limit normally exists.  For example, all the perturbative computations in quantum electrodynamics or in the  $SU(3)_c \times SU(2)_L \times U(1)_Y$ Standard Model of the particles interactions  are based precisely on the existence of this limit for the appropriate ratios of functional integrations.
In any path-integral expression one must  specify the so-called overall normalization, i.e. the choice of the functional integration which appears in  the  denominator, because different normalizations generally give rise to  different results. The path-integrals in which the normalization is not specified are not well defined.

 The limit procedure which has been mentioned  in Remark 2.1 ensures the validity of the following  two properties.

\item{(P1)} {\it Linearity.} If, in a given quantum field theory,  the functional  integration region $R$  is the union of two disjoint parts, $R = R_1 \cup R_2$, then  the path-integral over $R$ is the sum of the path-integrals over $R_1$ and over $R_2$.

\item{(P2)}  {\it Translation invariance}. Suppose that, in a given quantum field theory, any field configuration
$\phi (x)$ can be written as
$$
 \phi (x)=  \phi_0(x) + \psi (x) \; ,
$$
where $ \phi_0(x)$ is  fixed  and  the variable $\psi (x)$ can fluctuate. When the action $S[\phi ]$ is a quadratic function of the field variables,  the functional integration is invariant  [20] under  translation
$$
\left  \langle X(\phi ) \right  \rangle \equiv {\int D\phi  \; e^{i S[ \phi ]} \, X( \phi ) \over 
\int D\phi  \; e^{i S[ \phi ]} }
= {\int D\psi \; e^{i S[ { \phi_0} + \psi ]} \, X( {\phi_0} + \psi) \over
\int D\phi  \; e^{i S[ \phi  ]} }\; .
$$

\noindent The basic properties (P1) and (P2)  can also be understood as defining relations  because all our  functional integral computations  are based  precisely on these two properties exclusively. For instance, properties (P1) and (P2) have been used to write equation (1.4).

Each gauge orbit $A$  can be represented by a field configuration which admits a  \v Cech-de Rham representation, i.e.  a representative of the class $A$ can be described  by a collection of local variables which, in a good covering $\{ {\cal U}_a \}$ of $M$, are given by
$$
A \leftrightarrow  \left ( v^a , \lambda^{a b} , n^{a b c } \right ) \; .
\eqno(2.1)
$$
$v^a $ denotes a  1-form  locally defined in the open set ${\cal U}_a$; $\lambda^{a b}$ represents a 0-form in the intersection ${\cal U}_a \cap {\cal U}_b $  such that $v^a - v^b = d \lambda^{a b}$,  and  the integer $n^{a b c} $ is defined  in  ${\cal U}_a \cap {\cal U}_b \cap {\cal U}_c $ with the property  $\lambda^{ab} + \lambda^{bc}  + \lambda^{ca} = n^{abc}$.  In our notations, a particular representative element of the  DB class, which appears on the left-hand-side of the arrow $\leftrightarrow$, is  described by the collection of  \v Cech-de Rham field components that are shown on the right-hand-side of $\leftrightarrow$.
In particular a representative of a class $ \omega   \in \Omega^1(M) / \Omega^1_{\hbox{\petit Z}} (M)$  fulfills
$$
\omega \leftrightarrow  \left ( \omega^a , 0 , 0 \right ) \; ,
\eqno(2.2)
$$
where $\{ \omega^a \}$ are the restrictions in the open sets $\{ {\cal U}_a \}$ of a 1-form on $M$ also denoted by $\omega $.  {\it Vice versa}, given a 1-form $\alpha \in \Omega^1(M)$,  the DB class associated with $\alpha $   has   \v Cech-de Rham representation
$$
\alpha \leftrightarrow  \left ( \alpha^a , 0 , 0 \right ) \; ,
\eqno(2.3)
$$
where $\{ \alpha^a \}$ are the restrictions of  $\alpha \in \Omega^1(M) $ in the open sets $\{ {\cal U}_a \}$. Note that if $ \omega   \in \Omega^1(M) / \Omega^1_{\hbox{\petit Z}} (M)$  then $\int \omega * \omega  = \int \omega \wedge d \omega $ mod {\d Z}.

In the $U(1)$ Chern-Simons field theory, a typical path-integral computation ---that will appear below--- takes the form
$$
  \left \langle \! \!  { \left \langle   e^{2 \pi i \int \omega * \alpha }   \right \rangle } \! \!  \right \rangle \equiv   {\int D\omega \; e^{2 \pi i k \int \omega * \omega } \; e^{ 2 \pi i \int \omega * \alpha } \over \int D\omega \; e^{2 \pi i k \int \omega * \omega }}  \; ,
\eqno(2.4)
$$
where  $ \omega   \in \Omega^2_{\hbox{\petit Z}}(M)^* $ represents the integration variable, whereas  $\alpha$ can be interpreted as a given classical external source. Let us first consider  the case in which  $\alpha$ is the  DB class   associated with a 1-form  $\alpha \in \Omega^1(M) $.   Let  $ \alpha^\prime $  be the DB class such that
$$
 \alpha^\prime \leftrightarrow \left ( {1\over 2k } \alpha^a , 0 , 0 \right ) \; .
\eqno(2.5)
$$
One can put
$$
\omega = -  \alpha^\prime  + \widetilde \omega \; ,
\eqno(2.6)
$$
where  $  \alpha^\prime  $ is fixed and the variable   $\widetilde \omega \in \Omega^2_{\hbox{\petit Z}} (M)^*$  can fluctuate. Since $D\omega = D\widetilde \omega $ (property (P2)) and [5]
$$
k \int \omega * \omega + \int \omega * \alpha = k \int \widetilde \omega * \widetilde\omega \, - k \int {\alpha^\prime} * {\alpha^\prime} \; ,
\eqno(2.7)
$$
one finds
$$
 \left \langle \! \!  { \left \langle  e^{2 \pi i \int \omega * \alpha } \right  \rangle} \! \! \right \rangle   = e^{- 2 \pi i k \int \alpha^\prime * \alpha^\prime }\; \; { \int D \widetilde \omega \;  e^{2 \pi i k \int \widetilde \omega * \widetilde \omega} \over
\int D \omega \; e^{2 \pi i k \int  \omega * \omega} }  =  e^{- 2 \pi i k \int \alpha^\prime * \alpha^\prime } \; .
\eqno(2.8)
$$
Because $\alpha $ is  globally defined in $M$, one finally obtains [5]
$$
e^{- 2 \pi i k \int \alpha^\prime * \alpha^\prime} = e^{- (2 \pi i / 4 k) \int \alpha \wedge d\alpha } \; .
\eqno(2.9)
$$
This procedure can also be applied when $\alpha $ is a 1-current. In particular, for each oriented knot $C$ which belongs to a 3-ball $\cal B$ inside $M$, one can find a Seifert surface $\Sigma \subset M$ such that $C = \der \Sigma $. This equation can be written in terms of currents: $j_C = d \alpha_\Sigma$, where $j_C$ is the 2-current of the knot $C$ and $\alpha_\Sigma$ is the 1-current of $\Sigma$.
The 1-current $\alpha_\Sigma $ can be understood as  distributional limit of  1-forms in $M$. By construction one has
$ \int \omega \wedge d \alpha_\Sigma  = \int \omega \wedge j_C = \oint_C \omega $.

The DB class  $\eta_C \in H^1_D(M)^*$   defined  by
$$
\eta_C \leftrightarrow (\alpha^a_\Sigma , 0 , 0 )
\eqno(2.10)
$$
only depends on the knot $C$. Then
for any $ \omega   \in  \Omega^2_{\hbox{\petit Z}} (M)^*$ one has
$$
  \int \omega * \eta_C   = \int\omega \wedge d \alpha_\Sigma  = \oint_C \omega \qquad \hbox{mod {\d Z}} \; .
\eqno(2.11)
$$
For a  two components oriented link $C_1 \cup C_2 \subset {\cal B} \subset M $, the value of the linking number of $C_1$ and $C_2$ is given by
$$
\ell k (C_1, C_2) =
\int \alpha_{\Sigma_1} \wedge d \alpha_{\Sigma_2} \; ,
$$
with  $C_1 = \der \Sigma_1 $ and  $C_2 = \der \Sigma_2 $.  For a single oriented  framed knot $C \subset {\cal B}\subset M$, the integral $ \int \alpha_{\Sigma} \wedge d \alpha_{\Sigma}$ represents the self-linking number of $C = \der \Sigma $ which is defined to be the linking number of $C$ and its framing $C_{\rm f} $, $\int \alpha_{\Sigma} \wedge d \alpha_{\Sigma} \equiv  \ell k (C, C_{\rm f})=\int \alpha_{\Sigma} \wedge d \alpha_{\Sigma_{\rm f}}  $.

Consider now equation (2.8) in the case   $\alpha = \eta_L = \sum_j q_j \eta_{C_j} $ where $\{ C_j \}$ are the components of a framed oriented colored link $L = C_1 \cup \cdots \cup C_n \subset {\cal B} \subset M$ and $q_j $ denotes the color of $C_j$; in the DB formalism each color (or charge) $q_j$ must assume [5] integer values.   One obtains
$$
\left  \langle \! \! \! {\left \langle  e^{2 \pi i \sum_{j=1}^n q_j \oint_{C_j} \omega  } \right  \rangle} \! \! \! \right \rangle    =   e^{- (2 \pi i / 4 k ) \sum_{ij=1}^n q_i q_j \hbox{\d L}_{ij}} \; ,
\eqno(2.12)
$$
where the integers $\hbox{\d L}_{ij}$ are the matrix elements of the   linking matrix associated with $L$.  The result (2.12) can also be obtained by taking $\alpha \in \Omega^1(M)$ and considering  the  $\alpha \rightarrow \eta_L $ limit in  equation (2.8).

 Equation (2.12) also gives the complete solution [5] of the $U(1)$ Chern-Simons quantum field theory defined in $M = S^3$ because, in this case, any link belongs to a 3-ball.

The concluding  remarks of this section concern some general properties [5] of the expectation values in the abelian Chern-Simons theory.

\noindent {\bf Remark 2.2.} Let the colored oriented and framed  link  $ L^\prime = L \cup U \subset M $ be the union of the link $L$ with the unknot $U$. If $U$ belongs to a 3-ball which is disjoint from $L$, and $U$ has trivial framing ---i.e. its framing $U_{\rm f}$ satisfies $\ell k (U , U_{\rm f}) =0$--- then the expectation value of the holonomy associated with $L^\prime $ is equal to the expectation value of the holonomy associated with $L$.
Indeed the expectation value of the holonomy associated with $L^\prime$ is the product [5] of the expectation values associated with $L$ and with $U$ (this feature can be understood as the topological version of the cluster property of ordinary quantum field theories), and the expectation value of the holonomy associated with the unknot $U$ is  equal to the expectation value of the identity.

\vskip 0.6 truecm
\centerline {\includegraphics[width=3.5in]{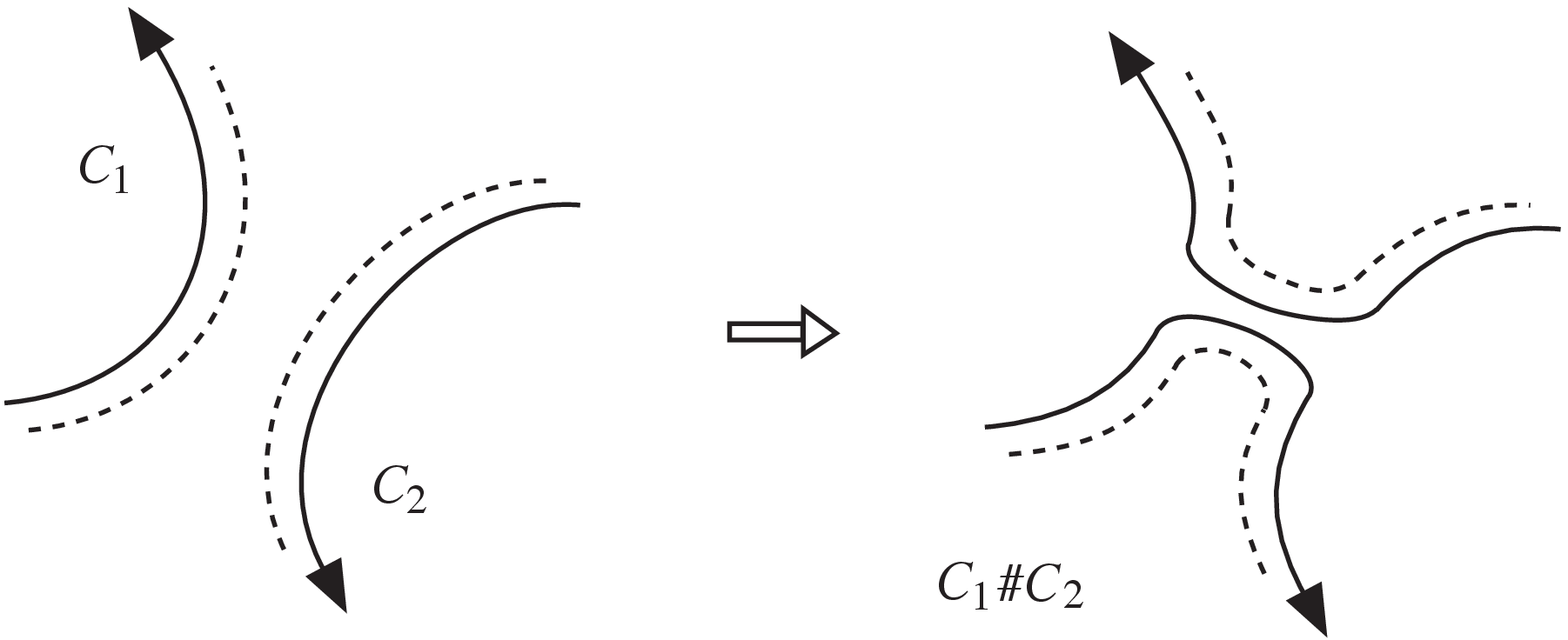}}
\vskip 0.3 truecm
\centerline {{Figure 2.1.} {Sum of knots.}}
\vskip 0.4 truecm

\noindent {\bf Definition 2.1.} {\it Let  $C_1$ and $C_2$ be two oriented (and possibly framed) knots in $M$.  By joining  $C_1$ and $C_2$ in the way shown in } Figure~2.1, {\it one obtains the knot $C_1 \# C_2$, that is called the (band connected) sum of $C_1$ and $C_2$.  The dashed lines in} Figure~2.1 {\it refer to the framings; by construction, the framing $(C_1 \# C_2)_{\rm f} $ of $C_1 \# C_2$ is just the sum of the framings $C_{1{\rm f}} \# C_{2{\rm f}} $. When $M = S^3$ ---which is of particular interest in the Dehn surgery presentation  in $S^3$ of  generic 3-manifold--- one has
$$
\ell k   ( (C_1 \# C_2)_{\rm f} , C_1 \# C_2  ) = {\ell} k ( C_{1{\rm f}} , C_1 ) + {\ell} k (  C_{2{\rm f}} , C_2 ) + 2 \, {\ell} k ( C_1 , C_2 ) \; .
\eqno(2.13)
$$
The knot $C_2$ with modified orientation  is indicated by $-C_2$; one finds }
$$
\ell k   ( (C_1 \# (-C_2))_{\rm f} , C_1 \# (-C_2)  ) = {\ell} k ( C_{1{\rm f}} , C_1 ) + {\ell} k (  C_{2{\rm f}} , C_2 ) - 2 \, {\ell} k ( C_1 , C_2 ) \; .
\eqno(2.14)
$$

\vskip 0.3 truecm

\noindent For colored knots $C_1$ and $C_2$, the sum $C_1 \# C_2$ is well defined when $C_1$ and $C_2$ have the same color. Since the linking number can be interpreted as an intersection product for which $\#$ plays the role of the standard sum, equations (2.13) and (2.14) also have a natural homological interpretation.
The sum of knots  enters the Kirby calculus [21].

\noindent  {\bf Remark 2.3.}   If the knots $C_1$ and $C_2$ have the same color, the expectation value of the holonomy associated with the link $L = C_1 \cup C_2 \cup C_3 \cup \cdots \cup C_n$ is equal to the expectation value of the holonomy associated with $L ^\prime = C_1 \# C_2 \cup C_3 \cup \cdots \cup C_n$. Indeed
the expectation values of the link holonomies are invariant under the addition in the link of a (trivially-framed) unknot which belongs to a 3-ball ${\cal B} \subset M$ (Remark 2.2); on the other hand, the band connected sum of two knots can be obtained by means of the introduction of  a trivially-framed unknot ---belonging to a 3-ball--- which coincides with the boundary of the band (precisely like in the case of the Kirby calculus [21]).

\vskip 1.2 truecm

\noindent {\sect 3. Path-integral partition function}

\vskip 0.4 truecm

In this section, the path-integral computation of the normalized partition function will be presented.  Following the general method introduced in [5],  we first  select ---as origin   $\widehat A_\gamma $---  a distributional class $\eta_\gamma \in H^1_D(M)^*$ which is canonically associated with a framed knot that represents   $\gamma \in H_1(M)$.    With this choice of $\widehat A_\gamma $, the value of the partition function  can be expressed by means of a sum of expectation values of knot holomomies.

Similarly to the homology group $H_1(M)$ ---that can equivalently be computed by using for instance the singular homology or the \v Cech homology---   the quadratic form $Q$ entering equation (1.13) can be determined by using different techniques.  The method  that we describe is essentially based on knot theory arguments; in Sect.3.1 we also mention an alternative procedure  which concerns flat (smooth) connections.  Both methods give  the same result.

\vskip 0.9 truecm

\noindent {\sect 3.1. Torsion knots and flat connections}

\vskip 0.4 truecm

Each element $\gamma \in H_1(M)$ can be represented by a oriented knot,   denoted by the same symbol $\gamma \subset M$.  Let $\eta_{\gamma} \in H^1_D(M)^*$ be the distributional  class [5] which is canonically associated with the knot $\gamma \subset M$.  Similarly to the case of the gauge orbit of a $U(1)$ gauge field, $\eta_\gamma$ admits a  \v Cech-de Rham representation
$$
\eta_\gamma \leftrightarrow  \left ( V_\gamma^a , \Lambda_\gamma^{a b} , N_\gamma^{a b c } \right ) \; ,
\eqno(3.1)
$$
in which $d V_\gamma^a $ is the  restriction in ${\cal U}_a$ of the 2-current of the knot $\gamma$.  If the knot $\gamma $ is homologically trivial,   one can find a representative of the class $\eta_\gamma$ with $\Lambda_\gamma^{ab} =0 $ and $N_\gamma^{abc} =0$, in agreement with equation (2.10). Whereas when the knot $\gamma $ does not represent the trivial element of $H_1(M)$, the components $\Lambda_\gamma^{a b} $ and $ N_\gamma^{a b c } $ are nontrivial.

Any  class $A\in H^1_D(M)^*$, which is associated with the $U(1)$ principal bundle that is labelled by $\gamma \in H_1(M)$, can be written [5] as
$$
A = \widehat A_\gamma + \omega = \eta_\gamma + \omega \; ,
\eqno(3.2)
$$
with $ \omega   \in \Omega^2_{\hbox{\petit Z}} (M)^*$. As already stated after equation (2.12),  instead of considering directly distributional configurations, one could start   with a  smooth DB class $\widehat A_\gamma \in H^1_D(M)$ and then   take the distributional  $\widehat A_\gamma \rightarrow \eta_\gamma$ limit; in both  cases one finds the same results.    One has
$$
 \int A * A =  \int \left ( \omega * \omega + 2  \, \omega * \eta_\gamma +  \eta_\gamma * \eta_\gamma \right ) \; .
\eqno(3.3)
$$
In order to give a well defined meaning to $\int \eta_\gamma * \eta_\gamma$, one can  introduce a framing for $\gamma$  (the specific procedure will be discussed in Section~3.2).   For now we note that, as a consequence of the definition [5] of the framing method, the introduction of  a framing for $\gamma$ has the effect of trivializing (from the Deligne-Beilinson  point of view) the star product $\eta_\gamma * \eta_\gamma$,   that is
$$
 \int  \eta_\gamma * \eta_\gamma =0 \qquad \hbox{mod {\d Z}}\; .
 \eqno(3.4)
$$
Therefore, for framed $\gamma$ one finds
$$
 { \int D \omega \; e^{i S[\, \widehat A_{\gamma} + \omega ] } \over \int D \omega \;  e^{i S[ \omega ] } } =
  { \int D \omega \; e^{i 2 \pi k \int \omega* \omega } e^{i 4 \pi  k \int \omega* \eta_\gamma} \over \int D \omega \;  e^{i S[ \omega ] } }     \; .
\eqno(3.5)
$$
Let us consider the integral $\int \omega * \eta_\gamma$; since $ \omega   \in  \Omega^2_{\hbox{\petit Z}} (M)^*$,  a representative of  the product $\omega * \eta_\gamma $  has \v Cech-de Rham structure
$$
\omega * \eta_\gamma \leftrightarrow \left ( \omega^a \wedge d V_\gamma^a , 0 , \cdots , 0 \right )  \; ,
\eqno(3.6)
$$
which does not depend on the nontrivial components $\Lambda_\gamma^{a b}$ and $N_\gamma^{a b  c}$ of the decomposition (3.1) of $\eta_\gamma$.   The crucial point now is that, for each  element $\gamma$ of the torsion group $T(M)$, one can find a representative 1-current $\alpha_\gamma $  that can be used ---in the integral $\int \omega * \eta_\gamma$--- in the  place of $\eta_\gamma$, i.e. $\forall \omega $
$$
 \int \omega * \eta_\gamma = \int \omega \wedge d \alpha_\gamma \qquad \hbox{mod {\d Z}}\; .
\eqno(3.7)
$$
Indeed, since $\gamma$ is a torsion knot, one can find a nonvanishing integer $p_\gamma$ such that  $ p_\gamma  \gamma = \Gamma = \der \Sigma$.  Consequently at the level of currents
$$
\alpha_\gamma = {1\over p_\gamma}  \alpha_\Sigma \; ,
\eqno(3.8)
$$
where $\alpha_\Sigma $ is the 1-current of the surface $\Sigma $.
Let $\widetilde \alpha_\gamma$ be the  class which is  associated with $\alpha_\gamma$,
$$
\widetilde \alpha_\gamma \leftrightarrow \left ( {1\over p_\gamma} \alpha^a_\Sigma , 0 , 0 \right ) \; ;
\eqno(3.9)
$$
then, by taking into account equation (3.6),  one has
$$
 \int \omega * \eta_\gamma = \int \omega \wedge d \alpha_\gamma = \int \omega *  \widetilde \alpha_\gamma \qquad \hbox{mod {\d Z}}\; .
\eqno(3.10)
$$
Thus one can compute the path-integral (3.5) by using the method that has been described in Section~2.
From equations (2.8) and (2.12) it follows that
$$\eqalign {
 { \int D \omega \; e^{i S[\, \widehat A_{\gamma} + \omega ] } \over \int D \omega \;  e^{i S[ \omega ] } } &=   \left \langle \! \!  { \left \langle   e^{4 \pi i k \int \omega * \widetilde \alpha_\gamma }   \right \rangle } \! \!  \right \rangle = e^{- 2 \pi i  k  \int \widetilde \alpha_\gamma * \widetilde \alpha_\gamma} \cr
 &= e^{- 2 \pi i  (k / p_\gamma^2)  \int \alpha_\Sigma \wedge d \alpha_\Sigma} =e^{- 2 \pi i k  N_\Gamma / p_\gamma^2}   \; ,   \cr }
\eqno(3.11)
$$
where the integer $N_\Gamma $ denotes  the self-linking number  of $\Gamma $ in $M$
$$
N_\Gamma = \ell k (\Gamma , \Gamma_{\rm f})\Big |_{M} \; .
\eqno(3.12)
$$
The self-linking number (3.12) is well defined because $\Gamma $ is homologically trivial and, as it is shown in the following section, one can actually choose $\Gamma$  to be a unknot inside a 3-ball $\cal B$ in the 3-manifold $M$.

\vskip 0.3 truecm

\noindent {\bf Definition 3.1.}  {\it  Let us  introduce the DB class $A^0_\gamma $ by means of the definition }
$$
A^0_\gamma = \eta_\gamma - \widetilde \alpha_\gamma \; .
\eqno(3.13)
$$

\vskip 0.3 truecm

\noindent From equations (3.1)-(3.10) it follows that a representative of the class $A^0_\gamma $ can be described by the  \v Cech-de Rham components
$$
A^0_\gamma \leftrightarrow  \left ( 0 , \widetilde \Lambda_\gamma^{a b} , N_\gamma^{a b c } \right ) \; .
\eqno(3.14)
$$
Since the first component of the representation (3.14)  is vanishing,   $A^0_{\gamma}$ corresponds to the gauge orbit of a flat connection. Equation (3.10) implies that, for any  $ \omega   \in \Omega^2_{\hbox{\petit Z}} (M)^*$, one has
$$
\int \omega * A^0_\gamma = 0 \qquad \hbox{mod {\d Z}} \; .
\eqno(3.15)
$$
Hence if, instead of $\eta_\gamma $, one takes  $A^0_\gamma $ as origin of the fibre over  $\gamma \in H_1 (M)$,  decomposition (3.2) reads
$$
A =  A^0_\gamma + \omega  \; ,
\eqno(3.16)
$$
with $ \omega   \in \Omega^2_{\hbox{\petit Z}} (M)^*$,  and  one finds
$$
 \int A * A =   \int  A^0_\gamma * A^0_\gamma + \int \omega * \omega  \qquad \hbox{mod {\d Z}}    \; .
 \eqno(3.17)
$$
Consequently the path-integral (3.11), that does not depend on the choice of the origin in the space of gauge orbits,    becomes
$$
 { \int D \omega \; e^{i S[\, \widehat A_{\gamma} + \omega ] } \over \int D \omega \;  e^{i S[ \omega ] } } = { \int D \omega \; e^{i S[ A^0_{\gamma} + \omega ] } \over \int D \omega \;  e^{i S[ \omega ] } }  = e^{ i S[ A^0_{\gamma}] } { \int D \omega \; e^{i S[  \omega ] } \over \int D \omega \;  e^{i S[ \omega ] } } = e^{ i S[ A^0_{\gamma}] }
  \; .
\eqno(3.18)
$$
This concludes the proof of Proposition~1.

\vskip 0.3 truecm

Equations (3.11) and (3.18) imply
$$
 e^{ i S[ A^0_{\gamma}]} = e^{- 2 \pi i k  N_\Gamma / p_\gamma^2} \; .
 \eqno(3.19)
 $$
In order to evaluate  the amplitude $e^{ i S[ A^0_{\gamma}]} = e^{- 2 \pi i k  N_\Gamma / p_\gamma^2}$ for each $\gamma \in T(M)$  and produce the expression of the quadratic form on the torsion group, it is convenient to introduce a surgery presentation of $M$.

 Before proceeding with the path-integral computation of $e^{ i S[ A^0_{\gamma}]} $, let us point out  the main features of the two different  choices ---$\eta_\gamma $ and $A^0_\gamma$--- of the origins for the fibres of the DB affine bundle $H_D^1(M)^*$. The  following $\hbox{\d R} / \hbox{\d Z} $-valued integrals clearly display the basic peculiarities of $\eta_\gamma $ and $A^0_\gamma$,
 $$\eqalign {
\int \eta_\gamma * \eta_\gamma = 0 \quad &, \quad \int \omega * \eta_\gamma \not= 0 \cr
\int A^0_\gamma * A^0_\gamma \not= 0 \quad &, \quad \int \omega * A^0_\gamma = 0 \; . \cr}
\eqno(3.20)
 $$
 For the  class $A^0_\gamma$ one can find  smooth representatives, whereas $\eta_\gamma $ is in essence  distributional; indeed  $\eta_\gamma $ has precisely been chosen  in order to trivialize, with a framed knot $\gamma$,  the star product $ \eta_\gamma * \eta_\gamma $.  Differently from $\eta_\gamma$, the class $A_\gamma^0$ can be taken as origin for $H_D^1(M)^*$ and for
 $H_D^1(M)$  as well.  Because $H^1(M)= 0$, for each $\gamma \in H_1(M)= T(M)$, $A_\gamma^0$ represents a canonical origin for $H_D^1$ exactly as the zero or vanishing class $\widehat A_0 =0 $ can be regarded as the canonical origin for the fibre over the element $0 \in H_1(M) \simeq H^2(M)$.
Let us recall that such a canonical origin $A^0_\gamma$ does not exist [5] on the fibres over the freely generated component $F(M)$ of $H_1(M)$.

From the knowledge of the \v Cech-de Rham representation (3.14), one could in principle compute the amplitude $e^{ i S[ A^0_{\gamma}]} $ directly by using the relation
$$
\int A_\gamma^0 * A^0_\gamma \; \; {\sot  {\hbox{\petit Z}} \over  = } \; \; \Big \langle N_\gamma^2 \;   \odot \, \widetilde \Lambda_\gamma^1 \; , \; \xi^0_3 \Big \rangle \; ,
\eqno(3.21)
$$
where $\widetilde \Lambda^1_\gamma $ and $N^2_\gamma$ denote the collections $\{ \widetilde \Lambda_\gamma^{ab} \}$ and $\{ N_\gamma^{abc} \}$ respectively, $\xi^0_3$ denotes the points generated by a polyhedral decomposition of the manifold $M$, the pairing $\langle ~,~ \rangle$ coincides with \v Cech  chain-cochain pairing and the product $\odot $ is precisely defined in Ref.[12]. Since $\Gamma = p_\gamma   \gamma $ is homologically trivial, one has $p_\gamma  N_\gamma^2 = \delta \, \Xi^1_\Gamma $ where $\delta $ denotes the \v Cech coboundary operator. Then [15]
$$
\int A_\gamma^0 * A^0_\gamma \; \; {\sot  {\hbox{\petit Z}} \over  = } \; \;  - {1\over p_\gamma}  \Big \langle \Xi^1_\Gamma \, , \, \tau^\gamma_1 \Big \rangle \; ,
\eqno(3.22)
$$
in which $\tau^\gamma_1$ represents a torsion cycle homologous to $\gamma $ in the \v Cech formalism. In particular, if the good covering $\{ {\cal U}_a \}$ of $M$ is also a good cover of $\gamma$, the polyhedral decomposition of $M$ can be chosen to give also a polyhedral decomposition of $\gamma$ in such a way that the collection $\tau^\gamma_1 $ is just the \v Cech cycle generated by this decomposition.  Finally, $\Big \langle \Xi^1_\Gamma \, , \, \tau^\gamma_1 \Big \rangle$ is the \v Cech equivalent of the intersection $\Gamma \cap  c $ where $c$ is a singular chain such that $\Gamma = b \, c$. These intersections [17,18] precisely define  the torsion quadratic form $Q$.

\vskip 0.9 truecm

\noindent {\sect 3.2 Surgery presentation}

\vskip 0.4 truecm

Each 3-manifold $M$ admits a integer surgery presentation in $S^3$; let the surgery instruction be described by the framed  link ${\cal L} = {\cal L}_1 \cup {\cal L}_2 \cdots \cup {\cal L}_m \subset S^3 $. The integer surgery coefficients $ \{ a_t \} $ (with $t = 1,2,..., m$) coincide with the self-linking numbers of the link components, i.e.  $ \{ a_t \} $ correspond to the diagonal elements of the linking matrix $\hbox {\d L}$. With the introduction of a orientation for $\cal L$, the linking matrix elements are given by $\hbox{\d L}_{ts} = \ell k ({\cal L}_t , {\cal L}_{s \rm f})$ where $ {\cal L}_{s \rm f}$ denotes the framing of $ {\cal L}_{s}$.  Let the homology group of $S^3 - {\cal L}$ be generated by  $\{ G_1 ,..., G_m \}$, each generator $G_t$ is a small circle linked with ${\cal L}_t$ and oriented in such a way that $\ell k (G_t , {\cal L}_t ) = 1$. The group $H_1 (M) $ has the presentation shown in equation (1.10);  a generic element $\gamma \in H_1 (M) $ can be written as
$$
\gamma = \sum_{i=1}^w n_i h_i   \;  ,
\eqno(3.23)
$$
where the generators $\{ h_i \}$ (with $i=1,2,...,w$) of $H_1 (M) $ can be expressed in terms of $\{ G_1 ,..., G_m \}$ as shown in equation (1.15) and each integer $n_i $ takes values in the residue class of integers modulo the torsion number $p_i\, $, i.e.  $n_i = 0,1,2,..., p_i -1$. The sum over the elements of $H_1(M)$ corresponds to a multiple sum over the integer coefficients $ \{ n_i \}$.  Equations (3.23) and (1.15) give
$$
\gamma = \sum_{i=1}^w \sum_{t =1}^m n_i B_{it} G_t \; .
\eqno(3.24)
$$

In the surgery presentation of a generic 3-manifold $M$, the ambient isotopy classes of (framed) links in $M $ are described by links in the complement of the surgery link $\cal L$ in $S^3$.  In particular, the knot   that represents the element $\gamma \in H_1(M)$ is described by a framed knot $\gamma \subset S^3 - \cal L$.

Let us introduce a correspondence between the composition law for the elements of $H_1(M)$ and a composition law for the framed knots in $S^3 - \cal L$ that represent these elements.

\vskip 0.3 truecm

\noindent {\bf Definition 3.2.}  {\it If ---as group elements--- $\gamma = \gamma_1 + \gamma_2$ where $\gamma_1 \not= \gamma_2$, then the corresponding framed knots are related as
$$
\gamma = \gamma_1 \# \gamma_2 \; .
\eqno(3.25)
$$
The group relation $\gamma= \gamma_1 - \gamma_2$ corresponds to $\gamma = \gamma_1 \# (- \gamma_2)$.
The group relation $2 \gamma = \gamma + \gamma$ corresponds to  the band connected sum of a knot with itself which is  defined to be
$$
2 \gamma = \gamma \# \gamma_{\rm f} \; ,
\eqno(3.26)
$$
where $\gamma_{\rm f}$ denotes the framing of $\gamma$.}

\vskip 0.3 truecm

Relations (3.25)  and (3.26) are consistent with the abelian composition law of the homology group;  only the framing choice of the knots needs to be discussed.
From the group relation $0 = \gamma - \gamma $ and equation (2.14) it follows that any  knot that represents the zero element of $H_1(M)$ must have trivial framing with respect to the sphere $S^3$ of the surgery presentation.
For each nontrivial torsion knot $\gamma $, one can find an integer $p_\gamma >1$ such that the knot $\Gamma = p_\gamma \gamma$ ---which is the sum of $p_\gamma$ copies of $\gamma$---  is homologically trivial; therefore, $\gamma$ also must have trivial framing with respect to $S^3$. To sum up, the consistency of Definition~3.2 requires that all the framed knots that represent the elements of $H_1(M)$ must have trivial framing with respect to the sphere of the surgery presentation.

The triviality condition ---with respect to surgery $S^3$--- for the framings  of the knots that represents the elements of $H_1(M)$ can also be obtained in the path-integral derivation of the surgery rules for the expectation values of the link holonomies; this issue will be discussed in a forthcoming article [22].

According to the Definition~3.2, the group relation (3.24) can be interpreted in terms of sum of knots; each framed knot $\gamma$ that represents an element of the homology group  can be obtained from the framed knots $\{ G_1 ,..., G_m \}$  by means of a finite sequence of band connected sum operations.  In order to obtain trivial framings for all the $\gamma $ knots, we shall  choose  the framings $\{ G_{1\rm f} ,..., G_{m\rm f} \}$ of $\{ G_1 ,..., G_m \}$ to be trivial with respect to the sphere $S^3$ of the surgery presentation
$$
\ell k (G_t , G_{t \rm f})  \Big |_{S^3} = 0 \quad , \quad \forall t =1,2,..,m \; .
\eqno(3.27)
$$

\vskip 3.9 truecm

\noindent {\sect 3.3 Linking numbers}

\vskip 0.4 truecm

The linking number $\ell k (C_1 , C_2) |_{S^3}$ of two knots $C_1\subset S^3 - {\cal L}$ and $C_2 \subset S^3 - {\cal L}$  ---computed with respect to $S^3$---  does not necessarily coincide with the linking number $\ell k (C_1 , C_2) |_{M}$
  of $C_1$ and $C_2$ which is possibly defined in $M$.   Let ${\cal L}_t $ (with fixed $t$) be one component of the surgery link $\cal L $ with integer surgery coefficient $a_t$.  In the surgery construction of $M$,  the interior $\vnt $ of a tubular neighborhood $V$ of ${\cal L}_t$ is removed from $S^3$; then $V$ is sewed  with $S^3 - \vnt $ by means of a boundary gluing homeomorphism $h : \der V \rightarrow \der (S^3 - \vnt )$.
   The framing ${\cal L}_{t \rm f}$ of  ${\cal L}_t$   is  isotopic in $S^3 - {\cal L}$ with the image $h (\mu )$ of  the meridian $\mu $ of the solid torus $V \subset M$.   Now the meridian $\mu \subset V \subset M$ has a canonical framing $\mu_{ \rm f}\subset V \subset M $ which  belongs to the boundary $\der V$;   $\mu_{\rm f}$  also represents a possible meridian for the solid torus $ V$ and  the two meridians $\mu $  and $\mu_{  \rm f}$  are parallel on the surface $\der V$.     The knot $h (\mu_{\rm f}) \subset S^3 - {\cal L}$  represents a possible framing for $h(\mu )$.  Since $h(\mu)$ is ambient isotopic with ${\cal L}_{t \rm f}$, the knot $h (\mu_{\rm f}) $  also defines a framing ${\cal L}_{t \rm f}^{\rm f}$ for ${\cal L}_{t \rm f}$ that  will be called  the canonical framing of ${\cal L}_{t \rm f}$.    Both $h(\mu) $ and $h (\mu_{ \rm f})$ belong to the boundary torus $\der (S^3 - \vnt )$ and, according to the surgery instructions, the linking number of $h(\mu) $ and $h(\mu_{ \rm f})$ is equal to  $ a_t$    with respect to the sphere $S^3$ of the surgery presentation
$$
\ell k ({\cal L}_{t \rm f} , {\cal L}_{t \rm f}^{\rm f} ) \Big |_{S^3}  = \ell k  ( h(\mu) ,  h(\mu_{\rm f}) ) \Big |_{S^3} = a_t \; .
\eqno(3.28)
$$
In the 3-manifold $M$, the knots $h(\mu) $ and $h (\mu_{\rm f})$ are ambient isotopic ---by construction--- with $\mu $ and $\mu_{\rm f}$ respectively. Any meridinal disc of $V$ whose boundary is $\mu$ does not intersect  $\mu_{\rm f}$. Therefore  if $\mu $ and $\mu_{\rm f}$ are transported inside   $V\subset M $,  $\mu $ and $\mu_{\rm f}$ are ambient isotopic in $M$ with two untied unknots which belong to  a 3-ball ${\cal B} \subset M$ and which are unlinked in the interior of $\cal B$. Consequently  one has
$$
\ell k ({\cal L}_{t \rm f} , {\cal L}_{t \rm f}^{\rm f} ) \Big |_{M}  = {\ell k } ( h(\mu) ,  h(\mu_{\rm f}) )  \Big |_{M} = {\ell k } ( \mu , \mu_{\rm f} ) \Big |_{M} = 0 \; .
\eqno(3.29)
$$
The property which is encoded in equations (3.28) and (3.29)  can also be expressed  in the following convenient form.

 Suppose  that a given framed knot in $M$ is described ---in the surgery presentation--- by a framed knot $C\subset S^3 - \cal L$  with framing $C_{\rm f}$ such that
$$
{\ell k } ( C ,  C_{\rm f} ) \Big |_{S^3} = 0 \; .
\eqno(3.30)
$$
Let us assume  that, in the sphere of the surgery presentation,  $C$  is ambient isotopic  ---as an unframed knot---  with the knot ${\cal L}_{t \rm f}$; in compact notations, this is denoted by $C \sim {\cal L}_{t \rm f}$.   Then one has
$$
{\ell k } ( C ,  C_{\rm f} ) \Big |_{M} = - \ell k ({\cal L}_{t \rm f} , {\cal L}_{t \rm f}^{\rm f})\Big |_{S^3} = - a_t \; .
\eqno(3.31)
$$
Indeed, under the action of $a_t$ right-handed twist homeomorphisms of a tubular
neighborhoods of $C$, equations (3.30) is transformed into equation (3.28) and equation (3.31) becomes equation (3.29).

In order to generalize equations (3.28) and (3.29), let us consider the sum ${\cal P} = {\cal L}_{t\rm f} \# {\cal L}_{ s\rm f}$, with  for instance   $t \not= s$.  In agreement with  equation (2.13), the canonical framing ${\cal P}_{\rm f}$ of $\cal P$
satisfies [19]
$$
\ell k ({\cal P}  , {\cal P}_{\rm f} ) \Big |_{S^3} = a_t + a_s + 2 \ell k ({\cal L}_t , {\cal L}_s) \; .
\eqno(3.32)
$$
According to the surgery construction, the knot ${\cal P} \subset S^3 - \cal L$  represents a framed  unknot  which belongs to a 3-ball in $M$ and
$$
\ell k ({\cal P} ,   {\cal P}_{\rm f} ) \Big |_{M} = 0 \; .
\eqno(3.33)
$$
Therefore equations (3.30) and (3.31) admit the following generalization.

\vskip 0.3 truecm

\noindent {\bf Lemma 3.1.} {\it Let $C \subset S^3 - \cal L$ be a framed knot with framing $C_{\rm f}$ such that
$$
\ell k (C , C_{\rm f}) \Big |_{S^3}= 0 \; .
\eqno(3.34)
$$
Suppose that  $C \sim {\cal L}^\#$, i.e. $C$ is ambient isotopic ---as a unframed knot--- with the knot ${\cal L}^\# \subset S^3 - \cal L$, in which  ${\cal L}^\#$ corresponds to a finite sequence  of band connected sums of the framed surgery link components $\{ {\cal L}_{t{\rm f}} \}$.  Let ${\cal L}^\#_{\rm f}$ be the canonical framing of ${\cal L}^\#$; then $C$ represents} [21] {\it a framed unknot in $M$ that belongs to a 3-ball inside the 3-manifold $M$   with self-linking number }
$$
{\ell k } ( C ,  C_{\rm f} )  \Big |_{M} =  -   {\ell k } ( {\cal L}^\# ,  {\cal L}^\#_{\rm f} )  \Big |_{S^3}\; .
\eqno(3.35)
$$

\vskip 0.9 truecm

\noindent {\sect 3.4 The quadratic form}

\vskip 0.4 truecm

Since a generic element $\gamma $ of $H_1(M)$ can be written as a linear combination (3.24) of the generators $\{ G_t  \}$, the element $\Gamma = p_\gamma \gamma $ reads
$$
\Gamma = p_\gamma  \gamma = \sum_{i=1}^w \sum_{t=1}^m n_i B_{it} \, p_\gamma \, G_t \; .
\eqno(3.36)
$$
In agreement with the Definition~3.2 and relation (3.36), the framed knot $\Gamma $ can be understood as a band connected sum of the knots $\{ G_t \}$.  Since each knot $G_t $ has trivial framing in $S^3$, the knot $\Gamma $ also has trivial framing in $S^3$,
$$
\ell k (\Gamma , \Gamma_{\rm f} ) \Big |_{S^3} = 0 \; .
\eqno(3.37)
$$
As $\Gamma $ is homologically trivial,  the knot $\Gamma$ is ambient isotopic  ---as an unframed knot--- with a knot $\Gamma^\#$ in $S^3 - \cal L$ which is  a band connected sum  of the surgery link  framing components $ \{ {\cal L}_{t\rm f} \}$.  Therefore, in order to find the  self-linking number of $\Gamma $ in $M$, one can use  the analogue of equations (3.34) and (3.35). We only need to determine  $ \Gamma^\#$.

The homology decomposition (1.14)  gives rise to the following relation
$$
{\cal L}_{t \rm f} \sim {\cal L}_{t \rm f}^{\#} =  \sum_{s=1}^m \,  \hbox{\d L}_{ts} \, G_s \; ;   
\eqno(3.38)
$$
i.e. in the sphere $S^3$ the knot ${\cal L}_{t \rm f} $  is ambient isotopic ---as an unframed knot---  with the knot ${\cal L}_{t \rm f}^{\#} $  which coincides with the band connected sum $\sum_{s=1}^m \,  \hbox{\d L}_{ts} \, G_s$. The inverse $\hbox{\d L}^{-1}$ of the linking matrix in $\hbox{\d R}^m $ has rational matrix elements, this means that
$$
\hbox{\d L}^{-1}_{ts} = {d_{ts} \over p} \; ,
\eqno(3.39)
$$
where $d_{st} \in \hbox{\d Z}$ and  $p = {\rm Det} \,  \hbox{\d L} = p_1 p_2 \cdots p_w > 0 $, where the torsion numbers $\{ p_i \}$ are the integers defined in equation (1.8). Therefore, for any $\gamma $ one can choose   $p_\gamma = p$.  Relation  (3.38) can also  be expressed  as
$$
p \, G_t \sim \left ( p \, G_t \right)^{\#} =
\sum_{s=1}^m \left ( p \, \hbox{\d L}^{-1}_{ts} \right ) {\cal L}_{s \rm f} =
\sum_{s=1}^m  d_{ts}\, {\cal L}_{s \rm f}\; .
\eqno(3.40)
$$
Consequently, with the choice $p_\gamma = p $, from equations (3.36) and (3.40) it follows that
$$
\Gamma  \sim \Gamma^{\#} =  \sum_{i=1}^w \sum_{t=1}^m \sum_{s=1}^m n_i B_{it} d_{ts} \, {\cal L}_{s \rm f}
\; ,
\eqno(3.41)
$$
where $n_i B_{it} d_{ts} $ are integers.
 By using  equations (2.13) and (2.14) recursively, one finds  that the canonical  framing $\Gamma_{\rm f}^\# $ of $\Gamma^\# $  is determined [21] by
$$\eqalign {
\ell k (\Gamma^\# , \Gamma_{\rm f}^\# )  \Big |_{S^3} &=
 \sum_{i=1}^w \sum_{t=1}^m \sum_{s=1}^m  \sum_{j=1}^w \sum_{r=1}^m \sum_{u=1}^m  n_i B_{it} d_{ts} \, n_j B_{jr} d_{ru} \, \hbox{\d L}_{s u}\cr
 &= p^2  \sum_{i=1}^w \sum_{j=1}^w n_i n_j  \left (  \sum_{t=1}^m \sum_{s=1}^m B_{it} B_{js} \hbox{\d L}^{-1}_{ts}  \right ) \; . \cr}
 \eqno(3.42)
 $$
Therefore, according to equation (3.35) of   Lemma~3.1,  one has
$$
N_\Gamma = \ell k (\Gamma , \Gamma_{\rm f} )  \Big |_M = - \ell k (\Gamma^\# , \Gamma_{\rm f}^\# )  \Big |_{S^3} =
- p^2  \sum_{i=1}^w \sum_{j=1}^w n_i n_j  Q_{ij}  \; ,
 \eqno(3.43)
 $$
where $Q_{ij}$ is shown in equation (1.17)
$$
Q_{ij} = \sum_{t,s=1}^m \, B_{it} \, B_{js} \, \hbox{\d L}^{-1}_{ts} \; .
$$
Then equation (3.19) takes the form
 $$
 e^{ i S[ A^0_{\gamma}]} = e^{ 2 \pi i k  \sum_{i j } n_i n_j Q_{ij}} \; ,
 \eqno(3.44)
 $$
 which coincides with equation (1.13).
Finally, according to  the definition (1.4), the explicit path-integral computation of the normalized partition function of the  $U(1)$ Chern-Simons theory gives
$$
Z_k(M)  = \sum_{n_1=0}^{p_1-1} \sum_{n_2=0}^{p_2-1}\cdots  \sum_{n_w=0}^{p_w-1}e^{ 2 \pi i k \sum_{ij} n_i n_j Q_{ij}  }  \; .
\eqno(3.45)
$$
This concludes the derivation of  expression (1.17).

The Reshetikhin-Turaev $U(1)$ surgery invariant $I_k(M)$ of the 3-manifold $M$, which admits a surgery presentation in $S^3$ with surgery link $\cal L$, is defined by [7,8,9,23]
$$
I_k(M) = (2k)^{-m/2} e ^{i \pi \sigma ({\cal L}) /4}\, \sum_{q_1=1}^{2k} \cdots \sum_{q_m=1}^{2k}
e^{- (2 \pi i / 4 k ) \sum_{ij=1}^m q_i q_j \hbox{\d L}_{ij}}
\; ,
\eqno(3.46)
$$
where $\sigma ({\cal L}) $ denotes the signature of the linking matrix $\hbox {\d L}$ which is associated with the surgery link $\cal L$ (which has $m$ components). The multiple sum ---which appears in expression (3.46)--- can be transformed  by means of  the  Deloup-Turaev reciprocity formula [24], which represents a generalization of the  reciprocity formula [25] for Gauss sums. The symmetric bilinear form on the lattice $W$ of Theorem~1  of the article [24] corresponds to the bilinear form which is defined by the linking matrix $\hbox {\d L}$, and the sum over the elements in the  dual lattice $W^{\bullet}$ is represented ---in our case--- by the sum over the elements of the homology group $H_1(M)$. According to equation (3) of Ref.~[24], one has
$$
Z_k(M) = \left ( p_1 p_2 \cdots p_w \right )^{1/2}\,  I_k (M) \; .
$$
In the particular example illustrated in the next section, the validity of the above relation  will also be verified by a direct application of the standard  [25] reciprocity formula for Gauss sums.

\vskip 4.2 truecm

\noindent {\sect 4. One  example}

\vskip 0.4 truecm

 Let us consider as illustrative example  the closed oriented 3-manifold $ M_{2,6}$ that corresponds to the  surgery link ${\cal L} \subset S^3$ shown in Figure~3.1.

 \vskip 0.7 truecm
\centerline {\includegraphics[width=1.4in]{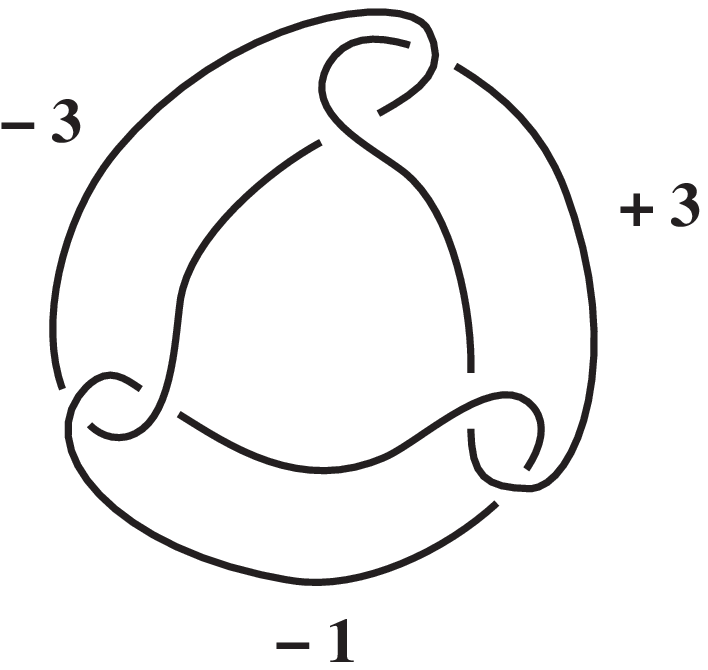}}
\vskip 0.3 truecm
\centerline {Figure 3.1. Surgery presentation in $S^3$ of the manifold $M_{2,6}$.}
\vskip 0.4 truecm

 \noindent The values of the surgery coefficients,  which are associated with the three link components ${\cal L } = {\cal L}_1 \cup {\cal L}_2 \cup {\cal L}_3 $, are $-3$,  $+3$ and $-1$ respectively.  One can introduce a orientation for $\cal L$ so that the linking matrix is given by
$$
\hbox{\d L} = \pmatrix { -3 & 1 & 1\cr 1 & 3 & 1 \cr 1 & 1 & -1 \cr } \; .
\eqno(4.1)
$$
 Let $\{ G_1 , G_2 , G_3 \} $ be the generators of the homology group of $S^3 - {\cal L}$;  $G_i$ (for $i=1,2,3$) is a small  circle linked with ${\cal L}_i$ and oriented in such a way that $\ell k (G_i , {\cal L}_i ) = +1$.   The generators  $G_1$, $G_2$ and $G_3$ have trivial framing with respect to the sphere $S^3$ of the surgery presentation,
$$
\ell k (G_t , G_{t \rm f})  \Big |_{S^3} = 0 \quad , \quad \hbox{for~~} =1,2,3 \; .
\eqno(4.2)
$$
The three conditions   $ \sum_{j}\hbox {\d L}_{ij} G_j = 0 $ (for $i=1,2,3$) that specify  $H_1(M_{2,6} )$ can be expressed in the form
$$\eqalign {
(a)& \quad G_3 = G_1 + G_2 \; , \cr
(b) & \quad 6 G_2 = 0 \; , \cr
(c) & \quad  2(G_1 - G_2) = 0 \; . \cr }
\eqno(4.3)
$$
Therefore   $ H_1(M_{2,6} ) = T(M_{2,6} )= \hbox{\d Z}_2 \oplus \hbox{\d Z}_6 $ in which one can take
$$\eqalign {
h_1 &= G_1 - G_2   \quad \hbox {as generator for~} \hbox{\d Z}_2\; , \cr
h_2 &= G_2  \hskip 1.5 truecm \hbox {as generator for~}  \hbox{\d Z}_6\; . \cr }
\eqno(4.4)
$$
Each element $\gamma \in H_1(M_{2,6} ) = \hbox{\d Z}_2 \oplus \hbox{\d Z}_6 $ can be written as
$$
\gamma = n_1 h_1 + n_2 h_2 \; ,   \quad \hbox{~with~~} n_1 = 0,1\hbox {~~and~~} n_2 = 0,1,2,3,4,5 \; .
\eqno(4.5)
$$
The inverse of the linking matrix is given by
$$
\hbox{\d L}^{-1}  =
{1\over 6}  \pmatrix { -2 & 1 & -1\cr 1 & 1 & 2 \cr -1 & 2 & -5 \cr }   \; ,
\eqno(4.6)
$$
and then, in the basis (4.4),  one finds
$$
Q = {1\over 6} \pmatrix {- 3 & 0 \cr 0 & 1 \cr} \; .
\eqno(4.7)
$$
The  normalized partition function is equal to
$$
Z_k(M_{2,6}) = \sum_{n_1 =0}^1 \sum_{n_2 =0}^5 e^{2 \pi i k (-3 n_1^2 + n_2^2) / 6 } \; .
\eqno(4.8)
$$
The Reshetikhin-Turaev invariant $I_k(M_{2,6})$ is given by
$$
I_k(M_{2,6}) = (2k)^{-3/2} e^{-i \pi /4 } \sum_{q_1=0}^{2k-1}\sum_{q_2=0}^{2k-1}\sum_{q_3=0}^{2k-1} e^{- (2 \pi i /4k) [ -3 q_1^2 + 2 q_1 q_2 + 3 q_2^2 + 2 q_2 q_3 - q_3^2 + 2 q_3 q_1] } \; .
\eqno(4.9)
$$
By means of the reciprocity formula [25] for the Gauss sums
$$
\sum_{n=0}^{|c|-1} e^{-{i\pi \over c} ( a n^2 +bn) } = \sqrt{| {c / a}| } \, e^{- {i \pi \over 4 ac } ( |ac | - b^2) }\,
\sum_{n=0}^{|a|-1} e^{{i\pi \over a} ( c n^2 +bn) } \; ,
\eqno(4.10)
$$
which is valid for integers $a$, $b$ and $c$ such that $ac \not= 0$ and $ac + b = \, $even,   one obtains
$$
\sum_{q_3=0}^{2k-1} e^{- (2 \pi i /4k) [ - q_3^2 +    2 q_3 (q_2 +   q_1)]} = \sqrt {2k}\;  e^{i\pi / 4} e^{- i( \pi / 2k ) (q_1 + q_2)^2} \; .
\eqno(4.11)
$$
Therefore
$$
I_k(M_{2,6}) = (2k)^{-1}  \sum_{q_1=0}^{2k-1}\sum_{q_2=0}^{2k-1} e^{- (2 \pi i /4k) [ -2 q_1^2 + 4 q_2^2 + 4 q_1 q_2 ] } \; .
\eqno(4.12)
$$
The reciprocity formula also gives
$$
 \sum_{q_1=0}^{2k-1}e^{ ( i \pi  /2k) [ 2 q_1^2 -  4 q_1 q_2 ] } = \sqrt {2k \over 2} \, e^{i \pi /4}\, e^{- i \pi q_2^2 / k} \sum_{n=0}^1e^{- i (\pi /2) [ 2 k n^2 -4nq_2 ] }\; .
 \eqno(4.13)
 $$
So one obtains
$$
I_k(M_{2,6}) = {1\over \sqrt 2} (2k)^{-1/2}  e^{i \pi /4} \sum_{n=0}^{1}  e^{- i \pi k n^2} \sum_{q_2=0}^{2k-1} e^{- (2 \pi i /4k) [   6 q_2^2 - 4 k n  q_2 ] } \; .
\eqno(4.14)
$$
Again the reciprocity formula produces
$$
\sum_{q_2=0}^{2k-1} e^{- (i \pi  /2k) [   6 q_2^2 - 4 k n  q_2 ] } = {\sqrt {2k}\over \sqrt {6}} \, e^{-i \pi /4} \, e^{i \pi k n^2 / 3} \sum_{m=1}^5 e^{i \pi (2k m^2 - 4k nm ) / 6}\; .
\eqno(4.15)
$$
Therefore
$$
I_k(M_{2,6}) = {1\over \sqrt {12  }}  \sum_{n=0}^{1}   \sum_{m=1}^{5} e^{ (2 \pi i k/6) [   -2  n^2 + m^2 - 2 m n ] } \; .
\eqno(4.16)
$$
Let us change variables and put
$$
n = n_1 \quad , \quad m = n_2 + n_1\; ,
\eqno(4.17)
$$
with $n_1 = 0 , 1$ and $n_2 = 0, 1,2,3,4,5$.
Expression (4.16) finally becomes
$$
I_k(M_{2,6}) = {1\over \sqrt {12}}  \sum_{n_1=0}^{1}   \sum_{n_2=1}^{5} e^{ 2 \pi i k (   -3 n_1^2 + n_2^2 ) / 6 } \; .
\eqno(4.18)
$$
Comparing equations (4.8) and (4.18), one finds
$$
Z_k (M_{2,6}) =  ( 2 \cdot  6  )^{1/2} \,  I_k (M_{2,6}) \; ,
\eqno(4.19)
$$
which is in agreement with equation (1.18).

\vskip 1.2 truecm

\noindent {\sect 5. Conclusions}

\vskip 0.4 truecm

In the $U(1)$ Chern-Simons theory the Deligne-Beilinson formalism sheds light on the fundamental role played by the flat connections in the computation of the topological invariants. In particular this formalism produces a path-integral  derivation of the abelian Reshetikhin-Turaev surgery invariant. In a forthcoming article [22] we will show in detail how to compute the link expectation values  in a generic 3-manifold by means of the Feynman functional integral; we will take into account both  the freely generated component and the torsion component of the homology group.

The Deligne-Beilinson approach to the Chern-Simons field theory also exists  [26] in a $(4n + 3)$-dimensional  closed manifold. Possible extensions of our results to these higher dimensional cases will be investigated.

\vskip 1.9 truecm

 \noindent {\bf Acknowledgments.}  We wish to thank Riccardo Benedetti for useful discussions.

\vskip 1.9 truecm

\noindent {\tmsm References}

\vskip 0.4 truecm

\item{[1]} A.S.~Schwarz, Lett. Math. Phys. 2 (1978) 247.

\medskip

\item{[2]} A.S.~Schwarz,  Commun. Math. Phys. 67 (1979) 1.

\medskip

\item{[3]} C.R.~Hagen, Ann. Phys. (N.Y.) 157 (1984) 342.

\medskip

\item{[4]} E.~Witten, Commun. Math. Phys. 121 (1989) 351.

\medskip

\item{[5]} E.~Guadagnini and F.~Thuillier, SIGMA 4 (2008) 078,  arXiv:0801.1445.

\medskip

\item{[6]} F.~Thuillier, J. Math. Phys. 50, 122301 (2009); arXiv:0901.2485.

\medskip

\item{[7]} N.Y.~Reshetikhin and V.G.~Turaev, Invent. Math. 103 (1991) 547.

\medskip

\item{[8]} H.~Murakami, T.~Ohtsuki, and M.~Okada, Osaka J. Math. 29 (1992) 545.

\medskip

\item{[9]} F.~Deloup, Math. Ann. 319 (2001) 759.

\medskip

\item{[10]} P.~Deligne, {\it Th\'eorie de Hodge II}, Publ. Math. I.H.E.S. n.40 (1971) 5.

\medskip

\item{[11]} A.A.~Beilinson, J. Soviet Math. 30 (1985) 2036.

\medskip

\item{[12]} J.L. Brylinski, {\it Loop spaces, characteristic classes and geometric quantization}, Progress in Mathematics, Vol. 107, Birkh?auser Boston, Inc., Boston, MA, 1993.

\medskip

\item{[13]} M.~Bauer, G.~Girardi, R.~Stora and F.~Thuillier, JHEP 0508 (2005) 027.

\medskip

\item{[14]} J. Cheeger and J. Simons, {\it Differential characters and geometric invariants}, Stony Brook Preprint 1973; reprinted in Lecture Notes in Mathematics 1167, Geometry and Topology Proc. 1983-84, Eds J. Alexander and J. Harer, Springer 1985.

\medskip

\item{[15]} J.L.~Koszul, {\it Travaux de S.S. Chern et J. Simons sur les classes caract\'eristiques}, S\'e\-mi\-nai\-re Bourbaki 26\`eme ann\'ee, n.440, (1973/74) 69.

\medskip

\item{[16]} M.~Kneser and P.~Puppe, Math. Zeitschr. 58 (1953) 376.

\medskip

\item{[17]} J.~Lannes and F.~Latour, {\it Forme quadratique d'enlacement et applications}, Ast\'erisque 26 (1975).

\medskip

\item{[18]} A.~Gramain, {\it Formes d'intersection et d'enlacement sur une vari\'et\'e}, M\'emoires de la Soci\'et\'e Math\'ematique de France, 48 (1976), p. 11-19.

\medskip

\item{[19]}  D.~Rolfsen, {\it Knots and Links}, AMS Chelsea Publishing, Providence, 2003.

\medskip

\item {[20]} D.~Elworthy and A.~Truman, {\it Feynman maps, Cameron-Martin formulae and anharmonic oscillators}, Ann. Inst. Henri Poincar\'e Phys. Th\'eor. 41 (1984) 115.

\medskip

\item{[21]}  R.~Kirby, Invent. Math. 45 (1978) 35.

\medskip

\item{[22]} E. Guadagnini and F. Thuillier, {\it Path integral and surgery rules in abelian Chern-Simons theory}, in preparation.

\medskip

\item{[23]} V.G.~Turaev, {\it Quantum invariants of knots and 3-manifolds}, de Gruyter Studies in Mathematics 18 (Berlin, 1994).

\medskip

\item{[24]} F.~Deloup and V.~Turaev, Journal of Pure and Applied Algebra 208 (2007) 153.

\medskip

\item{[25]} C.L.~Siegel, {\it \"Uber das quadratische Reziprozit\"atsgesetz algebraischen Zahlk\"orpern}, Nachr.  Acad. Wiss. G\"ottingen Math. Phys. Kl. 1 (1960) 1.

\medskip

\item{[26]} L.~Gallot, E.~Pilon and F.~Thuillier, {\it Higher dimensional abelian Chern-Simons theories and their link invariants}, arXix:1207.1270v1, to appear in J. Math. Phys.

\vfill\eject
\end
\bye